\documentclass[prl,reprint,
nolongbibliography, 
superscriptaddress,
aps,amsmath,amssymb,
]{revtex4-2}
\usepackage{graphicx}
\usepackage{xcolor}
\usepackage{braket}
\usepackage{bm}
\usepackage{hyperref}
\usepackage{ragged2e}  

\begin{document}


\title{Growing Einstein--Rosen Bridge with Multi-partite Entanglement}
\author{Takanori Anegawa}
\email{takanegawa@gmail.com}
\affiliation{Yonago College, National Institute of Technology, Yonago, Tottori 683-8502, Japan}
\author{Akihiro Miyata}
\email{a.miyata@maizuru-ct.ac.jp}
\affiliation{National Institute of Technology, Maizuru College, Kyoto 625-8511, Japan}
\affiliation{Yukawa Institute for Theoretical Physics, Kyoto University, Kyoto 606-8502, Japan}
\author{Shota Suzuki}
\email{chso25001@g.nihon-u.ac.jp}
\author{Kotaro Tamaoka}
\email{tamaoka.kotaro@nihon-u.ac.jp}
\affiliation{Department of Physics, College of Humanities and Sciences, Nihon University, Sakura-josui, Tokyo 156-8550, Japan}

\date{\today}

\begin{abstract}
Can quantum entanglement continue to grow even after the entanglement entropies of subsystems have saturated? We address this question using genuine multi-entropy, which probes
multi-partite entanglement structure beyond bipartite entanglement. In a two-sided BTZ
black hole, this quantity grows linearly with time after every subsystem
entropy of a fixed tripartition has saturated, extending the interval in which
entanglement detects Einstein--Rosen bridge growth. The result refines the
ER=EPR relation by identifying multi-partite structure invisible to ordinary
entanglement entropies. Some spin-chain models and the spatially partitioned Sachdev--Ye--Kitaev (SYK) model exhibit distinct
transients, while the late-time plateau values in holography, the SYK model, and Haar-random states share
a universal behaviour. 
\end{abstract}

\maketitle


\section{Introduction and Summary}
Can quantum entanglement continue to grow even after entanglement entropies have saturated? Entanglement growth and saturation have long served as central probes of nonequilibrium quantum dynamics~\cite{Calabrese:2005in,Calabrese:2007rg, Kim:2013etb, Liu:2013iza,Nahum:2016muy}. More broadly, tracking the delocalization of initially localized quantum information is central to characterizing relaxation and scrambling~\cite{Hosur:2015ylk,Mezei:2016wfz}. To address this question, it is useful to employ the Choi--Jamio{\l}kowski isomorphism, or channel--state map, which represents a unitary evolution $U(t)$ as a state obtained by acting on one half of a maximally entangled state (``EPR pairs'') $\ket{\Phi^+}$, 
\begin{align}
\ket{U(t)}_{LR}=(U(t)_L\otimes 1_R)\ket{\Phi^+}_{LR}\label{eq:channel}
\end{align}
on input ($L$) and output ($R$) Hilbert
spaces. Mutual informations between subsystems and suitable linear combinations thereof provide diagnostics of the scrambling ability of the unitary dynamics~\cite{Hosur:2015ylk}.

Once the relevant subsystem entropies become stationary, however, these
diagnostics cannot resolve any further evolution. Entanglement entropy
quantifies entanglement across a bipartition of a pure state; even the full
set of such entropies need not determine its multi-partite structure.
Whether multi-partite entanglement continues to develop after
entropic saturation, and what that development tells us about the dynamics,
therefore remain important questions. See \cite{Berthiere:2024sio,Balasubramanian:2025jhq, Fujiki:2026qdt} for recent developments. 

The AdS/CFT correspondence~\cite{Maldacena:1997re} sharpens these questions in a geometric
form. The channel state \eqref{eq:channel} with UV regularization can be regarded as a time-evolved
thermofield double (TFD) state~\cite{Israel:1976ur,Takahashi:1996zn}, dual to a two-sided eternal black hole in AdS with an
elongating Einstein--Rosen (ER) bridge~\cite{Maldacena:2001kr}. This is a concrete realization of the
ER=EPR slogan~\cite{Maldacena:2013xja} that connects entanglement in the boundary quantum mechanical systems and spacetime connectivity in the bulk~\cite{Ryu:2006bv, Hubeny:2007xt}. In particular, early linear
entropy growth reflects the stretching of extremal surfaces through the
elongating ER bridge~\cite{Hartman:2013qma}. Yet these entropies can saturate while the
bridge continues to grow. This tension motivated Susskind's observation that
``entanglement is not enough''~\cite{Susskind:2014moa}, and holographic complexity~\cite{Susskind:2014rva} has been widely studied as a probe of the subsequent interior dynamics.

In this work, we explicitly show that ordinary entropy saturation does not exhaust
the entanglement description of wormhole growth. To this end, we use multi-entropy~\cite{Gadde:2022cqi,Penington:2022dhr} and, to isolate its genuinely multi-partite entanglement contribution, genuine multi-entropy~\cite{Iizuka:2025ioc}. The definitions of these quantities will be reviewed in the following section. Remarkably, holographic multi-entropy continues to evolve and capture the growth of the ER bridge even after every subsystem entropy of a fixed tripartition has
saturated. 

Our findings sharpen the ER=EPR paradigm by showing that the entanglement associated with the growing ER bridge cannot be represented solely by independent EPR pairs, shared among the three parties. This suggests that describing
a sufficiently long, semiclassical wormhole requires extensive multi-partite entanglement
beyond the EPR pairs present at the initial time. 

We also examine spin chains and the spatially partitioned Sachdev--Ye--Kitaev (SYK) model~\cite{Sachdev:1992fk,KitaevTalks,Sachdev:2015efa,Maldacena:2016hyu}. Their
genuine multi-entropies exhibit model-dependent dynamics, including periodic behaviour ({\it i.e.}, quantum revival) absent from the semiclassical gravity result. Nevertheless, the holographic late-time value has the same leading dependence on subsystem sizes as a Haar-random state, and the SYK plateau closely approaches its finite-dimensional counterpart. The comparison identifies common late-time organization despite distinct transient behavior. We also give a simple three-qubit model where genuine multi-entropy increases while every bipartite entanglement spectrum stays fixed.

\section{Review and a simple example}
Here we briefly review (genuine) multi-entropy and present a
three-qubit example in which genuine multi-entropy varies while all bipartite
entanglement spectra remain fixed, anticipating the holographic result below.

Although multi-entropy is defined for an arbitrary number $q$ of parties, we
restrict ourselves to the tri-partite case, $q=3$, throughout this paper. See \cite{Gadde:2022cqi, Penington:2022dhr,Iizuka:2025ioc} for further details. Consider a tri-partite pure state
\begin{align}
 \ket{\Psi_3}
 =
 \sum_{\alpha_A,\alpha_B,\alpha_C}
 \psi_{\alpha_A\alpha_B\alpha_C}
 \ket{\alpha_A}\otimes\ket{\alpha_B}\otimes\ket{\alpha_C}.
\end{align}
For an integer $n\geq2$, label its $n^2$ replicas by
$(i,j)\in\mathbb Z_n^2$. Then, the replica invariant and the corresponding
tri-partite R\'enyi multi-entropy are defined~\footnote{Here we follow the normalization in \cite{Penington:2022dhr, Iizuka:2025ioc}, for example.} as
\begin{align}
 Z_n^{(3)}
 ={}&
 \sum_{\{\alpha_X^{(i,j)}\}}
 \prod_{i,j\in\mathbb Z_n}
 \psi_{\alpha_A^{(i,j)}
       \alpha_B^{(i,j)}
       \alpha_C^{(i,j)}}
 \nonumber\\[-2pt]
 &\hspace{15mm}\times
 \bar{\psi}_{\alpha_A^{(i+1,j)}
             \alpha_B^{(i,j+1)}
             \alpha_C^{(i,j)}},
 \\
 S_n^{(3)}
 =&
 \frac{1}{n(1-n)}
 \log\frac{Z_n^{(3)}}{\bigl(Z_1^{(3)}\bigr)^{n^2}}.
\end{align}
All replica labels are understood modulo $n$, and $Z_1^{(3)}=1$ for a
normalized state. The apparently distinguished role of $C$ is conventional;
the resulting multi-entropy is symmetric under permutations of the parties.

The genuine multi-entropy at R\'enyi index $n$ is then defined by
\begin{align}
 \mathrm{GM}_n^{(3)}
 =
 S_n^{(3)}
 -\frac{1}{2}
 \bigl[S_n(A)+S_n(B)+S_n(C)\bigr],
 \label{eq:GM_definition}
\end{align}
where we subtract the $n$-th R\'enyi entropies $S_n(X)=\frac{1}{1-n}\log\operatorname{Tr}\rho_X^n$.
This subtraction makes $\mathrm{GM}_n^{(3)}$ vanish for tensor products of
pure bipartite states shared among $AB$, $BC$, and $CA$~\cite{Penington:2022dhr}. 

In what follows, we use the analytically continued $n\to1$ limit in gravity and set $n=2$ for finite-dimensional quantum systems. We will suppress the subscript $n$ for all quantities evaluated in the $n\to1$ limit.

As an illustrative example for the gravity result discussed later, consider
\begin{align}
 \ket{\psi(\theta)}
&=\cos\theta\,\ket{G}+\sin\theta\,\ket{W},
 \qquad
0\leq\theta\leq\frac{\pi}{2},
 \label{eq:toy}
\end{align}
where
\begin{align}
     \hspace{-4mm}\ket{G} &=\frac{\ket{000}+\sqrt{2}\ket{111}}{\sqrt{3}},\,
\ket{W}=\frac{\ket{001}+\ket{010}+\ket{100}}{\sqrt{3}}.
\end{align}
For every $\theta$, each one-qubit reduced density matrix has eigenvalues
$\{2/3,1/3\}$. Since $\ket{\psi(\theta)}$ is pure, the nonzero spectrum of
each two-qubit reduced density matrix coincides with that of the complementary
one-qubit subsystem. Hence all bipartite entanglement spectra are independent
of $\theta$. The multi-entropies of the individual states $\ket{G}$ and $\ket{W}$ were previously evaluated in \cite{Gadde:2022cqi}.

One can check~\footnote{See Supplemental Material} that the $n=2$ genuine multi-entropy grows continuously as the state varies from $|\psi(\theta_0=\tan^{-1}\sqrt{2})\rangle$ to $|\psi(\frac{\pi}{2})\rangle=\ket{W}$, the W-state, even though every bipartite entanglement
spectrum remains fixed. 
Interestingly, $\ket{\psi(\theta_0)}$ is locally unitarily
equivalent to $\ket{G}$, a generalized GHZ state. This provides a finite-dimensional analogue of the
separation between entropic saturation and multi-partite-entanglement growth
found holographically below. 

\section{Gravity}

\begin{figure*}[t]
 \includegraphics[width=\textwidth]{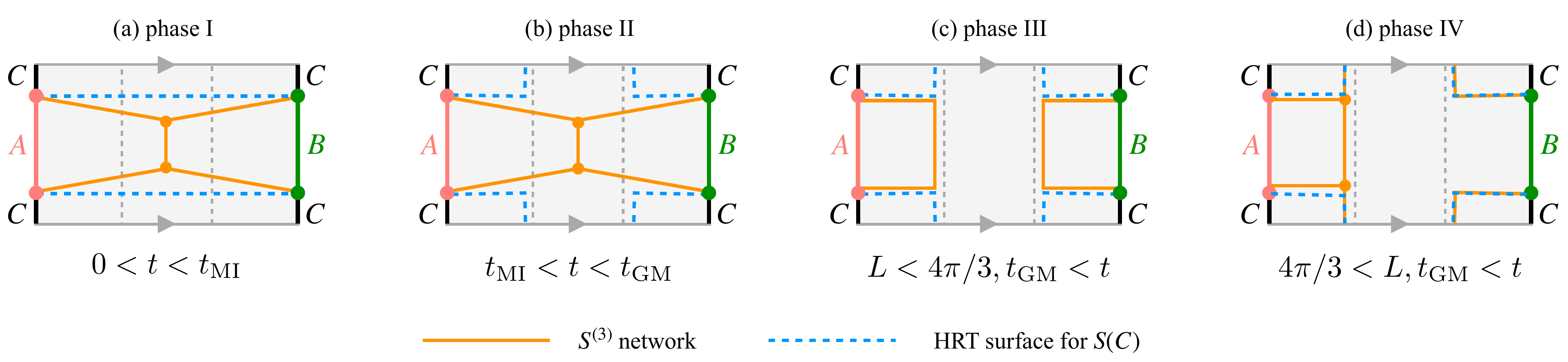}
 \caption{All four dynamical phases for $\pi<L<2\pi$. Orange curves are the
 multi-entropy network; blue dashed curves are the competing HRT surface for
 $S(C)=S(AB)$. The HRT surfaces for $S(A)$ and $S(B)$ are the ordinary interval
 caps in every panel and are omitted. The top and bottom edges are identified, so that the spatial direction forms a circle. The two gray vertical dashed lines indicate the horizons, and the intervening band represents the ER bridge. (a),(b) The same two-junction network survives the HRT
 transition. (c) It is replaced by two interval caps computing $S(A)$ and $S(B)$. (d) The network on one side forms a loop. Since $A$ and $B$ are chosen to have exactly the same size, the reflected configuration of (d) is degenerate with it. These are schematic diagrams, drawn in an exaggerated manner with the large-$r_+$ limit in mind.}
 \label{fig:phases}
\end{figure*}
First we investigate 2D holographic CFT. Let us consider a static BTZ black hole~\cite{Banados:1992wn},
\begin{align}
    ds^2
    =-(r^2-r_+^2)dt^2
    +\frac{dr^2}{r^2-r_+^2}
    +r^2d\phi^2 ,
    \label{eq:btz}
\end{align}
where $r_+$ denotes the horizon radius. We impose periodic boundary conditions on the angular coordinate, $\phi\sim\phi+2\pi$. The inverse Hawking temperature is then $\beta=\frac{2\pi}{r_+}$.
The maximally extended geometry describes a two-sided eternal black hole, with the exterior regions governed by the metric~\eqref{eq:btz}. As discussed above, its dual quantum state is the TFD state,
\begin{align}
    \ket{\mathrm{TFD}_{\beta}}
    =\frac{1}{\sqrt{Z(\beta)}}\sum_n
    e^{-\beta E_n/2}\ket{n}_L\ket{n^*}_R .
    \label{eq:tfd}
\end{align}
In the high-temperature limit, $\beta\to0$, or equivalently in the large-black-hole limit, $r_+\to\infty$, the TFD state approaches a maximally entangled state. Its time evolution can therefore be identified with the operator state associated with the time-evolution operator. Throughout this paper, we focus on this limit and the leading order in the large $r_+$ expansion. We are interested in unitary time-evolution of the entanglement structure associated with the TFD state. Following the convention commonly adopted in the literature, we use the doubled time-evolution operator $U(2t)_L=e^{i2H_Lt}$ acting on the TFD state, corresponding to a global quench from $H_L-H_R$ to $H_L+H_R$.

Throughout this paper, we focus on a symmetric configuration. The subsystems $A$ and $B$ are spatially aligned intervals of equal length $L$, located in the left and right systems, respectively. Their two complementary intervals are combined into the third subsystem $C$.

The holographic prescription computes the multi-entropy $S^{(3)}$ from an equal-tension extremal network that separates the three boundary regions~\cite{Gadde:2022cqi}. The network may contain trivalent junctions, at which the local force-balance condition in the spacelike plane fixes the opening angles to $2\pi/3$~\cite{Gadde:2022cqi,Harper:2024ker}. Crucially, the topology of this network is selected independently of that of the HRT surfaces computing the entanglement entropy. FIG.~\ref{fig:phases} displays both structures in each dynamical phase. All results below refer to the leading classical contribution in the high-temperature regime. In what follows, we focus on $L>\pi$, since for $L<\pi$ the subsystem $C$ occupies more than half of the total system, and bipartite entanglement becomes dominant. See also \cite{ Iizuka:2024pzm, Li:2025nxv, Anegawa:2025prn} for the same phenomenon in Haar-random states and holography. 

The competition between extremal surfaces for the holographic entanglement entropy in the current setup was studied in \cite{Goto:2021gve}. The entropies $S(A)$ and $S(B)$ remain constant~\footnote{In the large-$r_+$ limit, we need not consider geodesics that wind around the black hole, as they are always longer than the simple caps shown in the figures.}, while $S(C)=S(AB)$ is initially computed by two geodesics stretching across the wormhole. For $L>\pi$, these geodesics are eventually replaced by geodesics capping off the complementary intervals. The resulting mutual information is
\begin{equation}
    \frac{1}{2}I(A{:}B)_{L>\pi}
    =
    s_{\rm BH}
    \max\left\{L-2t,\,2L-2\pi\right\}
    ,
    \label{eq:mi-gravity}
\end{equation}
up to terms of order $r_+^0$. Here we defined the entropy density $s_{\rm BH}=\frac{r_+}{4G_{N}}$. After this transition, the entanglement entropies of the subsystems become stationary.

On the other hand, three admissible networks compete in the evaluation of the multi-entropy $S^{(3)}$~\footnote{See the Supplemental Material for details of the calculations}. (See FIG.~\ref{fig:phases}.) The connected network consists of two trivalent junctions, four external legs, and an internal branch connecting the two junctions through the black-hole interior. After the common UV-divergent contribution, which is irrelevant to the mutual information and genuine multi-entropy, is subtracted, the network has leading length $r_+(L+2t)$. In particular, the bulk-to-bulk internal branch contributes $r_+(L-2t)$. This negative time-dependent contribution is crucial for making the connected phase relatively long-lived compared with the corresponding HRT surfaces for the entanglement entropies. By contrast, the network formed by two disconnected interval caps has leading length $2r_+L$. The third network consists of a non-contractible loop together with a cap associated with the opposite complementary interval, and has leading length $r_+(4\pi-L)$. This phase is important for determining the maximum value of $\mathrm{GM}^{(3)}$, which agrees with the Haar-random-state result.

\begin{figure}[!t]
 \includegraphics[width=\columnwidth]{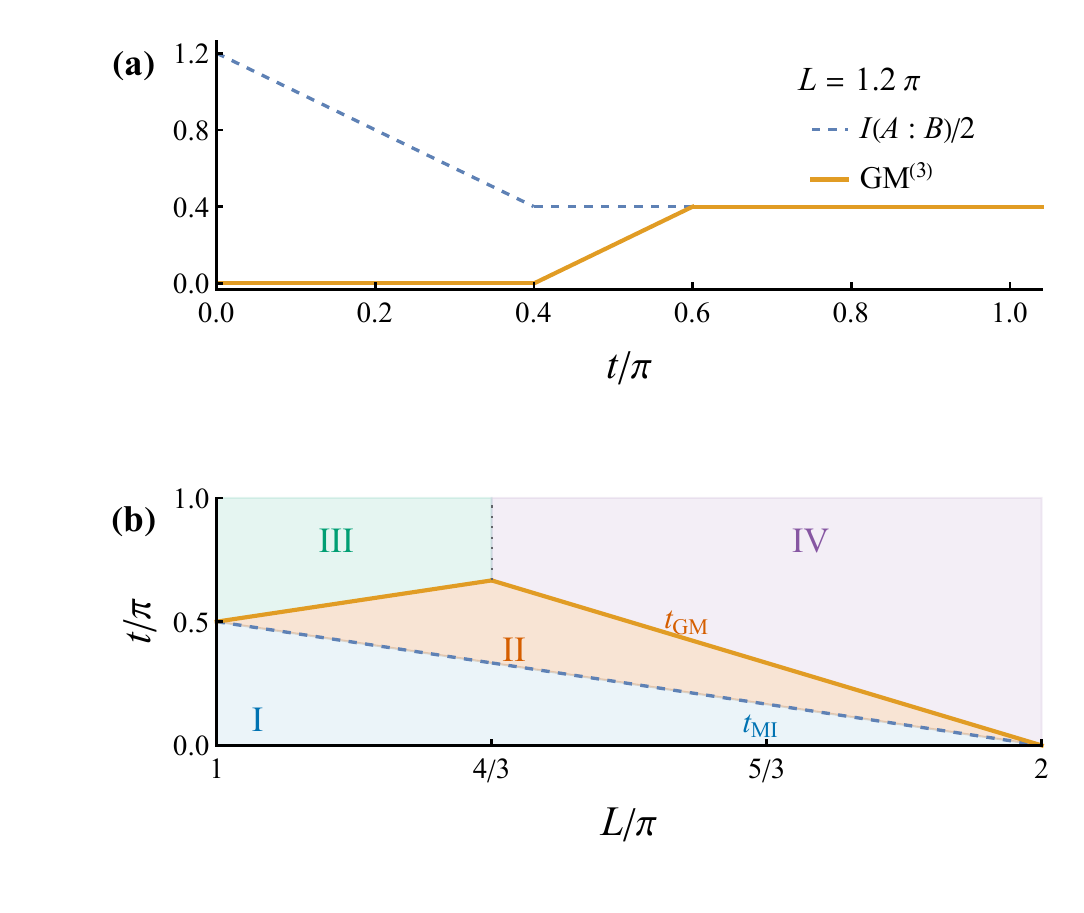}\vspace{-0.7cm}
 \caption{Leading high-temperature dynamics and phase structure. (a) $\textrm{GM}^{(3)}$ continues to grow even after the mutual information has reached its nonzero plateau. We set $L=1.2\pi$ and measure all quantities in units such that $\frac{\pi r_+}{4G_N}\equiv1$. (b) The phase diagram of $\textrm{GM}^{(3)}$ shows that this quantity extends the regime in which wormhole growth remains visible from phase I alone to phases I and II.
 }
 \label{fig:growth}
\end{figure}

The transitions in the entanglement entropy and multi-entropy therefore occur at different times $t_{\rm MI}$ and $t_{\rm GM}$,
\begin{equation}
 t_{\rm MI}=\pi-\frac L2,
 \qquad t_{\rm GM}=\min\left\{\frac L2,\,2\pi-L\right\}.
 \label{eq:times}
\end{equation}
Subtracting the bipartite contribution gives
\begin{align}
 &\hspace{-1mm}\textrm{GM}^{(3)}(A{:}B{:}C)_{L>\pi}\nonumber\\
 &\hspace{-1mm}=
 \left\{\begin{array}{@{}l@{\enspace}l@{}}
 0,&(0<t<t_{\rm MI}),\\
 2s_{\rm BH}(t-t_{\rm MI}),&(t_{\rm MI}<t<t_{\rm GM}),\\
 s_{\rm BH}\,\min\{2L-2\pi,2\pi-L\},&(t>t_{\rm GM}),
 \end{array}\right.
 \label{eq:gm-gravity}
\end{align}
up to terms of order $r_+^0$ at fixed $G_N$. 
and~\eqref{eq:gm-gravity} summarize four phases rather than three
(see FIG.~\ref{fig:growth}). The most important regime is phase II, in which all mutual informations have saturated, while the genuine multi-entropy continues to grow linearly in time. As is clear from FIG.~\ref{fig:phases}, this linear growth directly captures the growth of the ER bridge, which is invisible to entanglement entropies. The phase III and IV determines the late time plateau value. The maximal value is realized when the three subsystems $A$, $B$, and $C$ have equal size, namely $L=4\pi/3$, yielding
\begin{align}
    \max\left[\mathrm{GM}^{(3)}(A{:}B{:}C)\right]=\frac{1}{3}S_{\mathrm{BH}},
\end{align}
where $S_{\mathrm{BH}}$ is the Bekenstein--Hawking entropy. Remarkably, the late time value matches with the tri-partite Haar-random state with bound dimension $d=e^{S_{BH}}$. 

\section{Spin systems}
\begin{figure}[tb]
 \includegraphics[width=0.88\columnwidth]{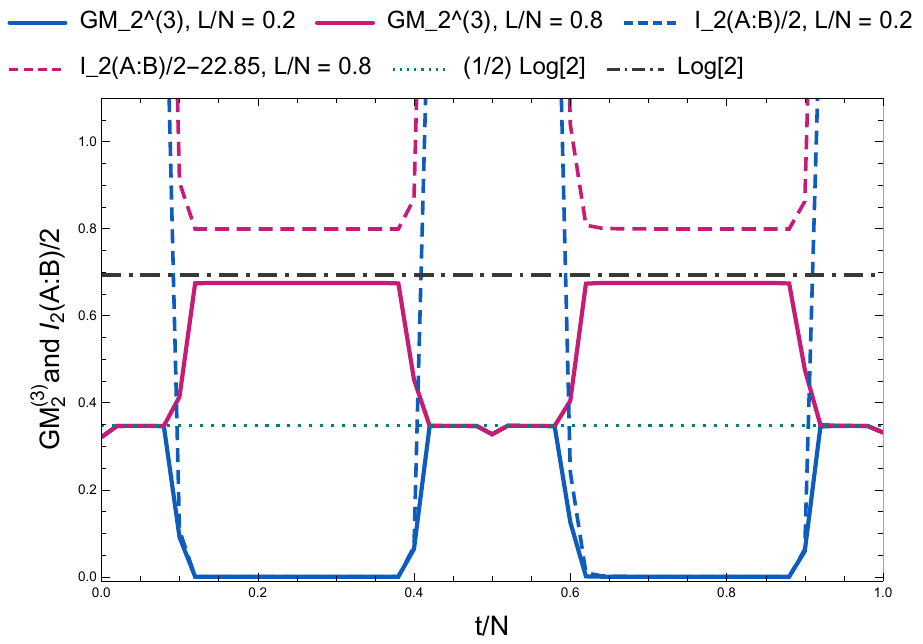}
 \caption{Time evolution of $\textrm{GM}_2^{(3)}$ and $I_2(A:B)/2$ in the TFD state of 1D hopping model($J = 0.5, N = 1000$, and $\beta = N/50$). The plots show the results for $L/N = 0.2$ and $0.8$, respectively. Also, since $I_2(A:B)/2$ becomes very large at $L/N = 0.8$, we have subtracted an appropriate constant for visual comparison only.}
 \label{fig:HopGM}
\end{figure}

\begin{figure}[tb]
 \includegraphics[width=0.88\columnwidth]{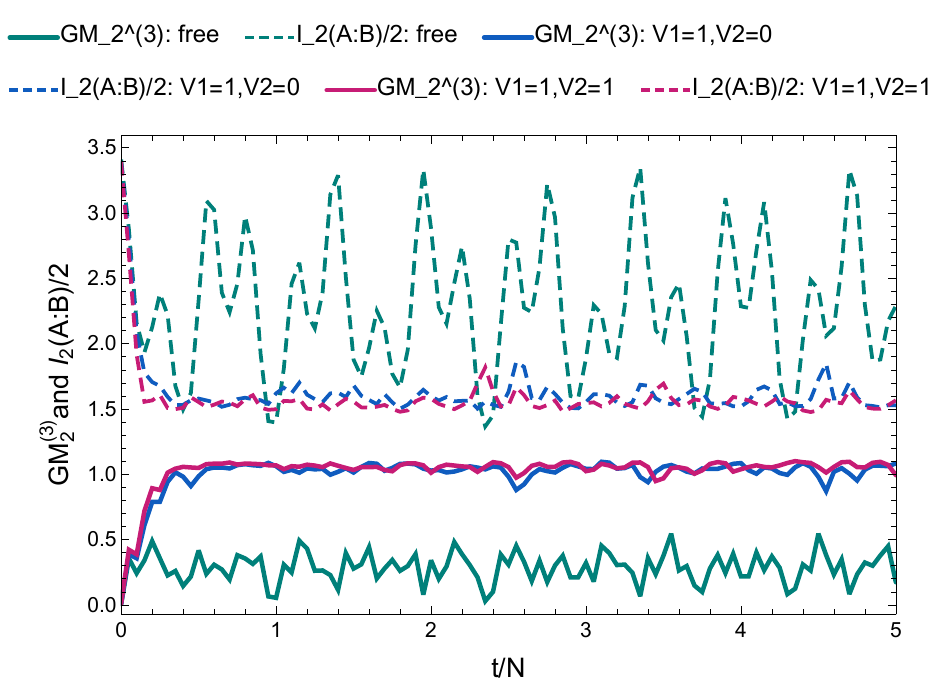}
 \caption{Time evolution of $\textrm{GM}_2^{(3)}$ and $I_2(A:B)/2$ for TFD state of a periodic fermion chain ($N=8,L=5,\beta/N=1/50$), computed by exact diagonalization for three interaction strengths: free($J=1,V_1=V_2=0$), integrable($J=V_1=1, V_2=0$) and non-integrable($J=V_1=V_2=1$)}
 \label{fig:GMwithint}
\end{figure}

Next, we consider $\textrm{GM}_2^{(3)}$ of TFD states of periodic chains with $N$ sites on each side. We take an $L$-site interval on the left as $A$, the symmetrically placed $L$-site interval on the right as $B$, and all remaining sites as $C$. We first consider the simple 1D hopping model
\begin{equation}
 H_{\rm h}=-J \sum_{i=1}^N (c_i^\dagger c_{i+1}+c_{i+1}^\dagger c_i),
 \label{eq:hopping}
\end{equation}
where $c_i$ denotes the Dirac fermion. FIG.~\ref{fig:HopGM} shows the time evolution of $\mathrm{GM}^{(3)}_2$ for the TFD state constructed in the energy eigenbasis, computed using the relation $S^{(3)}_2(A{:}B{:}C)=\frac{1}{2}S_R^{(2,2)}(A{:}B)+S_2(AB)$ \cite{Iizuka:2025elr} and correlator method \cite{Siva:2021cgo}. We work in the scaling regime where lattice effects are suppressed while the temperature is high compared with the finite-size scale, so that the system is effectively described by a free-fermion CFT.
Note that the $\textrm{GM}_2^{(3)}$ exhibits nearly binary behavior. 
And we observe a plateau of height $\frac{1}{2} \log 2$ and this is independent of size at the initial time. When $L < N/2$, the system transitions to zero value at a certain time, and when $L > N/2$, it transitions to a plateau of height $\log 2$ at a certain time~\footnote{This plateau value is justified by extrapolating from numerical calculations for finite values of $N$ to large $N$.}.
Most importantly, whenever $I_2(A:B)/2$ is on a plateau, $\textrm{GM}_2^{(3)}$ is also on a plateau. This synchronization totally differs from the holographic result.
An explanation of these results based on CFT 
 will be provided in \cite{Longversion}.

For small values of $N$, interactions can also be included. Specifically, consider the following interacting Hamiltonian,
\begin{align}
 H=H_{\rm h} + \sum_{k=1,2} \left[V_k \sum_i \left(n_i - \frac{1}{2}\right)\left(n_{i+k} - \frac{1}{2}\right)\right].
 \label{eq:interaction}
\end{align}
It is known that the system $V_1 \neq  0, V_2 = 0$ is interacting but is Bethe integrable. When $V_2 \neq 0$, it loses its integrability. We calculate the time evolution of $\textrm{GM}_2^{(3)}$ and $I_2(A:B)/2$ for these cases. We perform the calculation by explicitly diagonalizing. 
In the range where $2L\ll N$, $\textrm{GM}_2^{(3)}$ behave almost as constants. The interesting behavior arises in the intermediate region where $L/N \approx 0.6$. An example of a plot for this region is FIG.~\ref{fig:GMwithint}. $\textrm{GM}_2^{(3)}$ and $I_2(A:B)/2$ show a clear qualitative separation between the free fermion and the two interacting models, while the integrable and non-integrable curves are nearly indistinguishable in both plateau value and fluctuation amplitude. However, given the small system size ($N=8$) accessible to exact diagonalization, it remains unclear whether this reflects a genuine insensitivity of $\textrm{GM}_2^{(3)}$ to integrability breaking or merely a finite-size effect that would be resolved at large $N$. Distinguishing these possibilities is an interesting future direction. Notably, turning on interactions strongly suppresses the revival oscillations seen at this system size.

\section{Spatially Partitioned SYK model}

\begin{figure}[t]
\centering
\noindent\includegraphics[width=0.9\columnwidth]{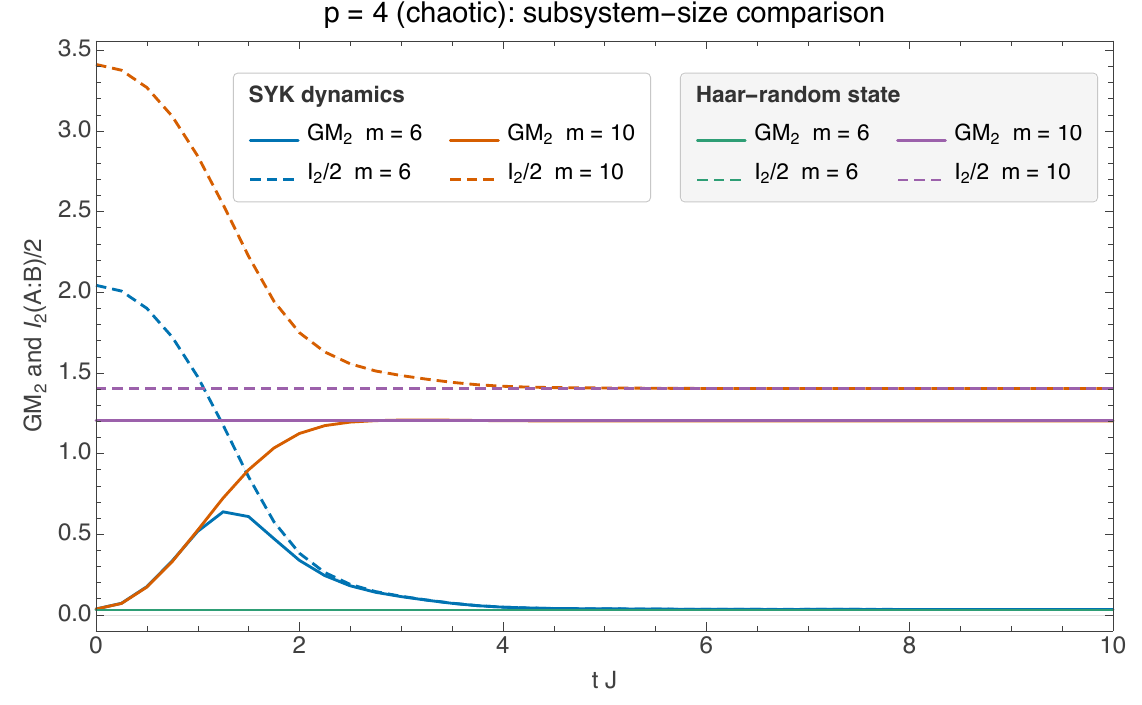}\llap{\makebox[\columnwidth][l]{\hspace{8mm}\raisebox{0.5\columnwidth}{\footnotesize\bfseries(a)}}}\par
\vspace{1.5mm}
\noindent\includegraphics[width=0.9\columnwidth]{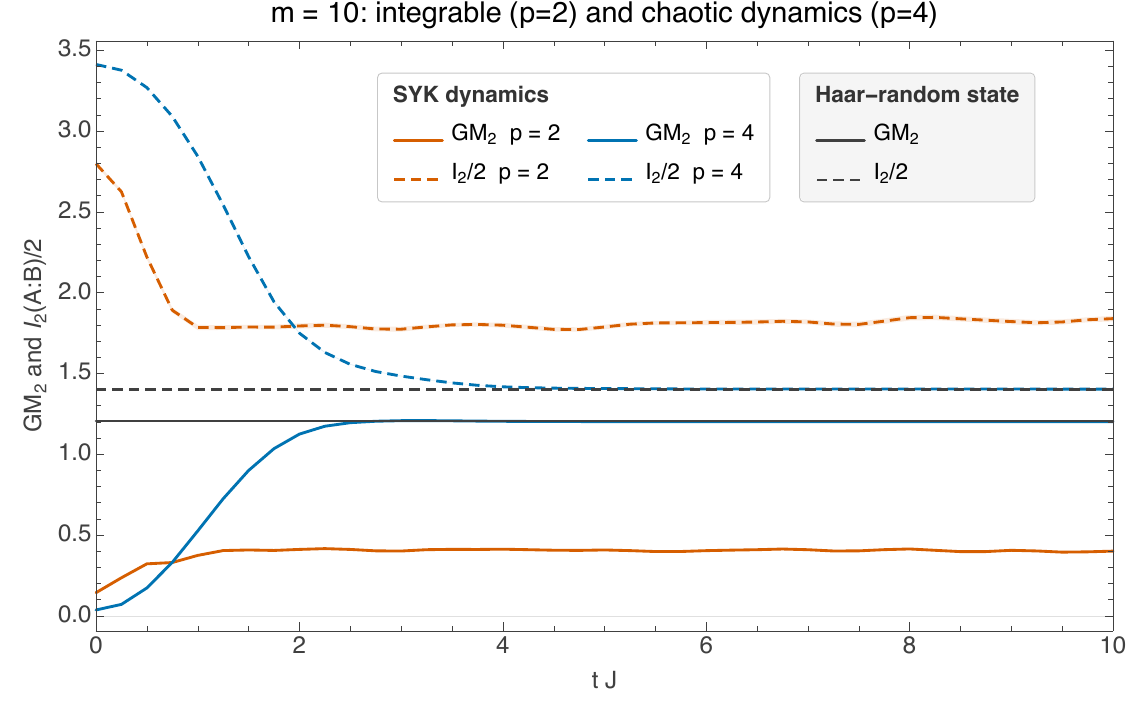}\llap{\makebox[\columnwidth][l]{\hspace{8mm}\raisebox{0.5\columnwidth}{\footnotesize\bfseries(b)}}}
\vspace{-2mm}
\caption{Time evolution of $\mathrm{GM}_2^{(3)}$ (solid) and $I_2(A{:}B)/2$ (dashed) in the SYK model for $N=16$ Majoranas per side and $\beta J=1$.
(a) $p=4$, $m=6,10$; (b) $m=10$, $p=2,4$.
SYK results are quenched averages over 64 disorder realizations for each $p$; Haar benchmarks are quenched averages over 256 unrestricted complex random pure states with $(d_A,d_B,d_C)=(2^{m/2},2^{m/2},2^{N-m})$.
Shading indicates one standard error.
Haar lines nearly overlap the $p=4$ plateaus; for $m=6$, the small late-time $\mathrm{GM}_2^{(3)}$ and $I_2/2$ also nearly coincide.}
\label{fig:syk_letter}
\end{figure}

We finally consider the $p$-body SYK model~\cite{Sachdev:1992fk,KitaevTalks,Sachdev:2015efa,Maldacena:2016hyu},
\begin{equation}
H_p=(i)^{\frac{p}{2}}\!\sum_{1\le i_1<\cdots<i_p\le N}\!J_{i_1\cdots i_p}\chi_{i_1}\cdots\chi_{i_p},
\label{eq:syk_hamiltonian}
\end{equation}
for $N$ Majoranas obeying $\{\chi_i,\chi_j\}=\delta_{ij}$.
The couplings $J_{i_1\cdots i_p}$ are independent real Gaussian random variables with zero mean and variance $\overline{J_{i_1\cdots i_p}^{\,2}}=(p-1)!J^2/N^{p-1}$.
We focus on $p=2,4$; the $p=2$ case is integrable, whereas $p=4$ is interacting and chaotic~\cite{KitaevTalks,Maldacena:2016hyu}.
We evolve the TFD state~\eqref{eq:tfd} with $e^{-it(H_L+H_R)}$, using identical disorder and $H_R=H_L^*$.
We pair the Majoranas into complex fermions and apply a Jordan--Wigner transformation, taking the first $m$ Majoranas (i.e., $m/2$ fermionic sites) on each side as $A$ and $B$, respectively, and the remainder as $C$.

FIG.~\ref{fig:syk_letter} shows exact-diagonalization results.
For $p=4$ and $m=10$, $\mathrm{GM}_2^{(3)}$ increases as $I_2$ decreases, then saturates close to the corresponding Haar-random benchmark.
Increasing $m$ from $6$ to $10$ makes the tripartition more balanced and raises the $\mathrm{GM}_2^{(3)}$ plateau, in agreement with the Haar benchmark (FIG.~\ref{fig:syk_letter}(a)).
The integrable $p=2$ model also reaches late-time plateaus, but for $m=10$ these remain distinct from the Haar benchmarks (FIG.~\ref{fig:syk_letter}(b)).

For $p=4$ and $m=6$, $\mathrm{GM}_2^{(3)}$ first rises and then decays to a small late-time value (FIG.~\ref{fig:syk_letter}(a)), so $\mathrm{GM}_2^{(3)}$ need not vary monotonically as $I_2$ decreases.

\section{Discussion}
We have shown that tri-partite entanglement continues to grow even after entanglement entropies and mutual informations have saturated in TFD state of 2D holographic CFT. Our results imply that extensive multi-partite entanglement is necessary to support a sufficiently long semiclassical ER bridge. Although the importance of multi-partite entanglement in AdS/CFT was already recognized~\cite{Umemoto:2019jlz, Akers:2019gcv, Iizuka:2025bcc}, our results make explicit its role in the emergence of spacetime in the deep bulk interior.

The comparison with the Markov gap $h(A:B)$ ~\cite{Akers:2019gcv,Hayden:2021gno} obtained from the reflected entropy~\cite{Dutta:2019gen} is particularly revealing. This quantity is computed holographically from the entanglement wedge cross section~\cite{Takayanagi:2017knl,Nguyen:2017yqw}. Owing to a discontinuous transition in the cross section, the Markov gap jumps to a constant value when the corresponding mutual information saturates. Therefore, at least as a probe of the growing black-hole interior, multi-entropy provides more fine-grained information.

We find that, both in gravity and in the spatially partitioned chaotic SYK model, the late-time value of genuine multi-entropy agrees with the corresponding Haar-random-state value. This agreement is particularly striking because these systems approach the universal plateau through qualitatively different dynamics. It suggests that sufficiently chaotic evolution may drive genuine multi-partite entanglement toward a universal typical-state value. Determining the precise conditions for this universality is an important direction for future work.

In the spin systems and SYK models studied here, we did not observe the asynchronous behavior found in holography. Although the use of different R\'enyi indices $n$ in the holographic and numerical calculations may partly account for this difference, our three-qubit example suggests that such behavior is not specific to a particular $n$. It is interesting to identify which many-body systems other than gravity exhibit this asynchronous behavior.

Although we have restricted our analysis to tri-partite entanglement, we expect multi-partite entanglement involving more parties to probe a broader regime of wormhole growth. It remains unclear, however, whether multi-entropies at general $q$ alone can probe the ER bridge at very late times, as holographic complexity is expected to do. Our results refine the statement of ER=EPR, but do not establish that ``multi-partite entanglement is enough.'' See also other entanglement-inspired approaches~\cite{Doi:2023zaf, Anegawa:2025tio, Fujiki:2026ucr} that probe the black hole interior. Clarifying this issue is an important future direction for understanding how spacetime emerges in holography.

\begin{acknowledgments}
\section{Acknowledgments}
We thank Norihiro Iizuka and Tadashi Takayanagi for fruitful discussions and useful comments on the draft. T.A.~is supported by JSPS KAKENHI Grant No.~26K17161. A.M.~is supported by Grant-in-Aid for JSPS Fellows 26KJ0186. K.T.~is supported by JSPS KAKENHI Grant No.~26K07072. We acknowledge OpenAI’s ChatGPT and Anthropic’s Claude for their assistance with coding for numerical calculations and manuscript editing.
All AI-assisted code and text were independently reviewed and verified by the authors, who take full responsibility for the content of the manuscript. 
\end{acknowledgments}

\bibliography{reference.bib}

@article{Banados:1992wn,
	archiveprefix = {arXiv},
	author = {Banados, Maximo and Teitelboim, Claudio and Zanelli, Jorge},
	doi = {10.1103/PhysRevLett.69.1849},
	eprint = {hep-th/9204099},
	journal = {Phys. Rev. Lett.},
	pages = {1849--1851},
	reportnumber = {PRINT-92-0151 (CHILE), IASSNS-HEP-92-29},
	title = {{The Black hole in three-dimensional space-time}},
	volume = {69},
	year = {1992}}

@article{Hartman:2013qma,
	archiveprefix = {arXiv},
	author = {Hartman, Thomas and Maldacena, Juan},
	doi = {10.1007/JHEP05(2013)014},
	eprint = {1303.1080},
	journal = {JHEP},
	pages = {014},
	primaryclass = {hep-th},
	title = {{Time Evolution of Entanglement Entropy from Black Hole Interiors}},
	volume = {05},
	year = {2013}}

@article{Maldacena:1997re,
	archiveprefix = {arXiv},
	author = {Maldacena, Juan Martin},
	doi = {10.4310/ATMP.1998.v2.n2.a1},
	eprint = {hep-th/9711200},
	journal = {Adv. Theor. Math. Phys.},
	pages = {231--252},
	reportnumber = {HUTP-97-A097, HUTP-98-A097},
	title = {{The Large N limit of superconformal field theories and supergravity}},
	volume = {2},
	year = {1998}}

@article{Ryu:2006bv,
	archiveprefix = {arXiv},
	author = {Ryu, Shinsei and Takayanagi, Tadashi},
	doi = {10.1103/PhysRevLett.96.181602},
	eprint = {hep-th/0603001},
	journal = {Phys. Rev. Lett.},
	pages = {181602},
	reportnumber = {NSF-KITP-06-11},
	title = {{Holographic derivation of entanglement entropy from AdS/CFT}},
	volume = {96},
	year = {2006}}

@article{Maldacena:2001kr,
	archiveprefix = {arXiv},
	author = {Maldacena, Juan Martin},
	doi = {10.1088/1126-6708/2003/04/021},
	eprint = {hep-th/0106112},
	journal = {JHEP},
	pages = {021},
	reportnumber = {NSF-ITP-01-59},
	title = {{Eternal black holes in anti-de Sitter}},
	volume = {04},
	year = {2003}}

@article{Hubeny:2007xt,
	archiveprefix = {arXiv},
	author = {Hubeny, Veronika E. and Rangamani, Mukund and Takayanagi, Tadashi},
	doi = {10.1088/1126-6708/2007/07/062},
	eprint = {0705.0016},
	journal = {JHEP},
	pages = {062},
	primaryclass = {hep-th},
	reportnumber = {DCPT-07-13, KUNS-2069},
	title = {{A Covariant holographic entanglement entropy proposal}},
	volume = {07},
	year = {2007}}

@article{Dutta:2019gen,
    author = "Dutta, Souvik and Faulkner, Thomas",
    title = "{A canonical purification for the entanglement wedge cross-section}",
    eprint = "1905.00577",
    archivePrefix = "arXiv",
    primaryClass = "hep-th",
    doi = "10.1007/JHEP03(2021)178",
    journal = "JHEP",
    volume = "03",
    pages = "178",
    year = "2021"
}

@article{Susskind:2014moa,
    author = "Susskind, Leonard",
    title = "{Entanglement is not enough}",
    eprint = "1411.0690",
    archivePrefix = "arXiv",
    primaryClass = "hep-th",
    doi = "10.1002/prop.201500095",
    journal = "Fortsch. Phys.",
    volume = "64",
    pages = "49--71",
    year = "2016"
}

@article{Hosur:2015ylk,
    author = "Hosur, Pavan and Qi, Xiao-Liang and Roberts, Daniel A. and Yoshida, Beni",
    title = "{Chaos in quantum channels}",
    eprint = "1511.04021",
    archivePrefix = "arXiv",
    primaryClass = "hep-th",
    reportNumber = "MIT-CTP-4733, MIT-CTP/4733",
    doi = "10.1007/JHEP02(2016)004",
    journal = "JHEP",
    volume = "02",
    pages = "004",
    year = "2016"
}

@article{Maldacena:2013xja,
    author = "Maldacena, Juan and Susskind, Leonard",
    title = "{Cool horizons for entangled black holes}",
    eprint = "1306.0533",
    archivePrefix = "arXiv",
    primaryClass = "hep-th",
    doi = "10.1002/prop.201300020",
    journal = "Fortsch. Phys.",
    volume = "61",
    pages = "781--811",
    year = "2013"
}

@article{Susskind:2014rva,
    author = "Susskind, Leonard",
    title = "{Computational Complexity and Black Hole Horizons}",
    eprint = "1403.5695",
    archivePrefix = "arXiv",
    primaryClass = "hep-th",
    doi = "10.1002/prop.201500092",
    journal = "Fortsch. Phys.",
    volume = "64",
    pages = "24--43",
    year = "2016",
    note = "[Addendum: Fortsch.Phys. 64, 44--48 (2016)]"
}

@article{Gadde:2022cqi,
    author = "Gadde, Abhijit and Krishna, Vineeth and Sharma, Trakshu",
    title = "{New multipartite entanglement measure and its holographic dual}",
    eprint = "2206.09723",
    archivePrefix = "arXiv",
    primaryClass = "hep-th",
    reportNumber = "TIFR/TH/22-34",
    doi = "10.1103/PhysRevD.106.126001",
    journal = "Phys. Rev. D",
    volume = "106",
    number = "12",
    pages = "126001",
    year = "2022"
}

@article{Penington:2022dhr,
    author = "Penington, Geoff and Walter, Michael and Witteveen, Freek",
    title = "{Fun with replicas: tripartitions in tensor networks and gravity}",
    eprint = "2211.16045",
    archivePrefix = "arXiv",
    primaryClass = "hep-th",
    doi = "10.1007/JHEP05(2023)008",
    journal = "JHEP",
    volume = "05",
    pages = "008",
    year = "2023"
}

@article{Iizuka:2025ioc,
    author = "Iizuka, Norihiro and Nishida, Mitsuhiro",
    title = "{Genuine multientropy and holography}",
    eprint = "2502.07995",
    archivePrefix = "arXiv",
    primaryClass = "hep-th",
    doi = "10.1103/714c-byxq",
    journal = "Phys. Rev. D",
    volume = "112",
    number = "2",
    pages = "026011",
    year = "2025"
}

@article{Harper:2024ker,
    author = "Harper, Jonathan and Takayanagi, Tadashi and Tsuda, Takashi",
    title = "{Multi-entropy at low Renyi index in 2d CFTs}",
    eprint = "2401.04236",
    archivePrefix = "arXiv",
    primaryClass = "hep-th",
    reportNumber = "YITP-24-02",
    doi = "10.21468/SciPostPhys.16.5.125",
    journal = "SciPost Phys.",
    volume = "16",
    number = "5",
    pages = "125",
    year = "2024"
}

@article{Li:2025nxv,
    journal = {},
    author = "Li, Zhi and Mori, Takato and Yoshida, Beni",
    title = "{Tripartite Haar random state has no bipartite entanglement}",
    eprint = "2502.04437",
    archivePrefix = "arXiv",
    primaryClass = "quant-ph",
    reportNumber = "YITP-25-15, RUP-25-8",
    month = "2",
    year = "2025"
}

@article{Anegawa:2025prn,
    author = "Anegawa, Takanori and Suzuki, Shota and Tamaoka, Kotaro",
    title = "{Black Holes as a Multipartite Entanglers: Multientropy in AdS3/CFT2}",
    eprint = "2512.21037",
    archivePrefix = "arXiv",
    primaryClass = "hep-th",
    doi = "10.1093/ptep/ptag047",
    journal = "PTEP",
    volume = "2026",
    number = "4",
    pages = "043B03",
    year = "2026"
}

@article{Israel:1976ur,
    author = "Israel, W.",
    title = "{Thermo field dynamics of black holes}",
    doi = "10.1016/0375-9601(76)90178-X",
    journal = "Phys. Lett. A",
    volume = "57",
    pages = "107--110",
    year = "1976"
}

@article{Takahashi:1996zn,
    author = "Takahashi, Y. and Umezawa, H.",
    title = "{Thermo field dynamics}",
    doi = "10.1142/S0217979296000817",
    journal = "Int. J. Mod. Phys. B",
    volume = "10",
    pages = "1755--1805",
    year = "1996"
}

@article{Calabrese:2005in,
    author = "Calabrese, Pasquale and Cardy, John L.",
    title = "{Evolution of entanglement entropy in one-dimensional systems}",
    eprint = "cond-mat/0503393",
    archivePrefix = "arXiv",
    doi = "10.1088/1742-5468/2005/04/P04010",
    journal = "J. Stat. Mech.",
    volume = "0504",
    pages = "P04010",
    year = "2005"
}

@article{Calabrese:2007rg,
    author = "Calabrese, Pasquale and Cardy, John",
    title = "{Quantum Quenches in Extended Systems}",
    eprint = "0704.1880",
    archivePrefix = "arXiv",
    primaryClass = "cond-mat.stat-mech",
    doi = "10.1088/1742-5468/2007/06/P06008",
    journal = "J. Stat. Mech.",
    volume = "0706",
    pages = "P06008",
    year = "2007"
}

@article{Liu:2013iza,
    author = "Liu, Hong and Suh, S. Josephine",
    title = "{Entanglement Tsunami: Universal Scaling in Holographic Thermalization}",
    eprint = "1305.7244",
    archivePrefix = "arXiv",
    primaryClass = "hep-th",
    reportNumber = "MIT-CTP/4475, MIT-CTP-4475",
    doi = "10.1103/PhysRevLett.112.011601",
    journal = "Phys. Rev. Lett.",
    volume = "112",
    pages = "011601",
    year = "2014"
}

@article{Kim:2013etb,
    author = "Kim, Hyungwon and Huse, David A.",
    title = "{Ballistic Spreading of Entanglement in a Diffusive Nonintegrable System}",
    eprint = "1306.4306",
    archivePrefix = "arXiv",
    primaryClass = "quant-ph",
    doi = "10.1103/PhysRevLett.111.127205",
    journal = "Phys. Rev. Lett.",
    volume = "111",
    number = "12",
    pages = "127205",
    year = "2013"
}

@article{Nahum:2016muy,
    author = "Nahum, Adam and Ruhman, Jonathan and Vijay, Sagar and Haah, Jeongwan",
    title = "{Quantum Entanglement Growth Under Random Unitary Dynamics}",
    eprint = "1608.06950",
    archivePrefix = "arXiv",
    primaryClass = "cond-mat.stat-mech",
    doi = "10.1103/PhysRevX.7.031016",
    journal = "Phys. Rev. X",
    volume = "7",
    number = "3",
    pages = "031016",
    year = "2017"
}

@article{Mezei:2016wfz,
    author = "Mezei, M{\'a}rk and Stanford, Douglas",
    title = "{On entanglement spreading in chaotic systems}",
    eprint = "1608.05101",
    archivePrefix = "arXiv",
    primaryClass = "hep-th",
    doi = "10.1007/JHEP05(2017)065",
    journal = "JHEP",
    volume = "05",
    pages = "065",
    year = "2017"
}

@article{Berthiere:2024sio,
    journal = {},
    author = "Berthiere, Cl{\'e}ment",
    title = "{Tripartite entanglement dynamics following a quantum quench}",
    eprint = "2408.12533",
    archivePrefix = "arXiv",
    primaryClass = "cond-mat.stat-mech",
    month = "8",
    year = "2024"
}

@article{Balasubramanian:2025jhq,
    author = "Balasubramanian, Vijay and Jiang, Hanzhi and Ross, Simon F.",
    title = "{Time evolution of multi-party entanglement signals}",
    eprint = "2511.16729",
    archivePrefix = "arXiv",
    primaryClass = "hep-th",
    doi = "10.1007/JHEP06(2026)055",
    journal = "JHEP",
    volume = "06",
    pages = "055",
    year = "2026"
}

@article{Sachdev:1992fk,
    author = "Sachdev, Subir and Ye, Jinwu",
    title = "{Gapless spin fluid ground state in a random, quantum Heisenberg magnet}",
    eprint = "cond-mat/9212030",
    archivePrefix = "arXiv",
    reportNumber = "PRINT-93-0077",
    doi = "10.1103/PhysRevLett.70.3339",
    journal = "Phys. Rev. Lett.",
    volume = "70",
    pages = "3339",
    year = "1993"
}

@misc{KitaevTalks,
  author = {A.~Kitaev},
  title = {A simple model of quantum holography.},
  howpublished = "\url{http://online.kitp.ucsb.edu/online/entangled15/kitaev/},\url{http://online.kitp.ucsb.edu/online/entangled15/kitaev2/}",
}

@article{Sachdev:2015efa,
    author = "Sachdev, Subir",
    title = "{Bekenstein-Hawking Entropy and Strange Metals}",
    eprint = "1506.05111",
    archivePrefix = "arXiv",
    primaryClass = "hep-th",
    doi = "10.1103/PhysRevX.5.041025",
    journal = "Phys. Rev. X",
    volume = "5",
    number = "4",
    pages = "041025",
    year = "2015"
}

@article{Maldacena:2016hyu,
    author = "Maldacena, Juan and Stanford, Douglas",
    title = "{Remarks on the Sachdev-Ye-Kitaev model}",
    eprint = "1604.07818",
    archivePrefix = "arXiv",
    primaryClass = "hep-th",
    doi = "10.1103/PhysRevD.94.106002",
    journal = "Phys. Rev. D",
    volume = "94",
    number = "10",
    pages = "106002",
    year = "2016"
}

@article{Takayanagi:2017knl,
    author = "Takayanagi, Tadashi and Umemoto, Koji",
    title = "{Entanglement of purification through holographic duality}",
    eprint = "1708.09393",
    archivePrefix = "arXiv",
    primaryClass = "hep-th",
    reportNumber = "YITP-17-89, IPMU17-0115",
    doi = "10.1038/s41567-018-0075-2",
    journal = "Nature Phys.",
    volume = "14",
    number = "6",
    pages = "573--577",
    year = "2018"
}

@article{Nguyen:2017yqw,
    author = "Nguyen, Phuc and Devakul, Trithep and Halbasch, Matthew G. and Zaletel, Michael P. and Swingle, Brian",
    title = "{Entanglement of purification: from spin chains to holography}",
    eprint = "1709.07424",
    archivePrefix = "arXiv",
    primaryClass = "hep-th",
    doi = "10.1007/JHEP01(2018)098",
    journal = "JHEP",
    volume = "01",
    pages = "098",
    year = "2018"
}

@article{Akers:2019gcv,
    author = "Akers, Chris and Rath, Pratik",
    title = "{Entanglement Wedge Cross Sections Require Tripartite Entanglement}",
    eprint = "1911.07852",
    archivePrefix = "arXiv",
    primaryClass = "hep-th",
    doi = "10.1007/JHEP04(2020)208",
    journal = "JHEP",
    volume = "04",
    pages = "208",
    year = "2020"
}

@article{Hayden:2021gno,
    author = "Hayden, Patrick and Parrikar, Onkar and Sorce, Jonathan",
    title = "{The Markov gap for geometric reflected entropy}",
    eprint = "2107.00009",
    archivePrefix = "arXiv",
    primaryClass = "hep-th",
    doi = "10.1007/JHEP10(2021)047",
    journal = "JHEP",
    volume = "10",
    pages = "047",
    year = "2021"
}

@article{Siva:2021cgo,
    author = "Siva, Karthik and Zou, Yijian and Soejima, Tomohiro and Mong, Roger S. K. and Zaletel, Michael P.",
    title = "{Universal tripartite entanglement signature of ungappable edge states}",
    eprint = "2110.11965",
    archivePrefix = "arXiv",
    primaryClass = "quant-ph",
    doi = "10.1103/PhysRevB.106.L041107",
    journal = "Phys. Rev. B",
    volume = "106",
    number = "4",
    pages = "L041107",
    year = "2022"
}

@unpublished{Longversion,
  author = {Anegawa, Takanori and Miyata, Akihiro and Suzuki, Shota and Tamaoka, Kotaro},
  title  = {},
  note   = {in preparation}
}

@article{Iizuka:2025bcc,
    journal = {},
    author = "Iizuka, Norihiro and Lin, Simon and Nishida, Mitsuhiro",
    title = "{Why many-partite entanglement is essential for holography}",
    eprint = "2504.01625",
    archivePrefix = "arXiv",
    primaryClass = "hep-th",
    month = "4",
    year = "2025"
}

@article{Umemoto:2019jlz,
    author = "Umemoto, Koji",
    title = "{Quantum and Classical Correlations Inside the Entanglement Wedge}",
    eprint = "1907.12555",
    archivePrefix = "arXiv",
    primaryClass = "hep-th",
    reportNumber = "YITP-19-72",
    doi = "10.1103/PhysRevD.100.126021",
    journal = "Phys. Rev. D",
    volume = "100",
    number = "12",
    pages = "126021",
    year = "2019"
}

@article{Goto:2021gve,
    author = "Goto, Kanato and Mollabashi, Ali and Nozaki, Masahiro and Tamaoka, Kotaro and Tan, Mao Tian",
    title = "{Information scrambling versus quantum revival through the lens of operator entanglement}",
    eprint = "2112.00802",
    archivePrefix = "arXiv",
    primaryClass = "hep-th",
    reportNumber = "RIKEN-iTHEMS-Report-21",
    doi = "10.1007/JHEP06(2022)100",
    journal = "JHEP",
    volume = "06",
    pages = "100",
    year = "2022"
}

@article{Iizuka:2024pzm,
    author = "Iizuka, Norihiro and Lin, Simon and Nishida, Mitsuhiro",
    title = "{Black hole multi-entropy curves {\textemdash} secret entanglement between Hawking particles}",
    eprint = "2412.07549",
    archivePrefix = "arXiv",
    primaryClass = "hep-th",
    doi = "10.1007/JHEP03(2025)037",
    journal = "JHEP",
    volume = "03",
    pages = "037",
    year = "2025"
}

@article{Iizuka:2025elr,
    author = "Iizuka, Norihiro and Miyata, Akihiro and Nishida, Mitsuhiro",
    title = "{Multipartite Markov gaps and entanglement wedge multiway cuts}",
    eprint = "2507.15262",
    archivePrefix = "arXiv",
    primaryClass = "hep-th",
    doi = "10.1007/JHEP10(2025)148",
    journal = "JHEP",
    volume = "10",
    pages = "148",
    year = "2025"
}

@article{Fujiki:2026qdt,
    journal = {},
    author = "Fujiki, Kosei and Tasuki, Kenya",
    title = "{Saddle Incompatibility Generates Genuine Multientropy in Heavy Local Quenches}",
    eprint = "2606.12526",
    archivePrefix = "arXiv",
    primaryClass = "hep-th",
    reportNumber = "YITP-26-65",
    month = "6",
    year = "2026"
}

@article{Doi:2023zaf,
    author = "Doi, Kazuki and Harper, Jonathan and Mollabashi, Ali and Takayanagi, Tadashi and Taki, Yusuke",
    title = "{Timelike entanglement entropy}",
    eprint = "2302.11695",
    archivePrefix = "arXiv",
    primaryClass = "hep-th",
    reportNumber = "YITP-23-22",
    doi = "10.1007/JHEP05(2023)052",
    journal = "JHEP",
    volume = "05",
    pages = "052",
    year = "2023"
}

@article{Fujiki:2026ucr,
    journal = {},
    author = "Fujiki, Kosei and Harper, Jonathan and Takayanagi, Tadashi and Zenoni, Nicol{\`o}",
    title = "{The Entanglement Wedge Polygon}",
    eprint = "2606.21081",
    archivePrefix = "arXiv",
    primaryClass = "hep-th",
    reportNumber = "YITP-26-76",
    month = "6",
    year = "2026"
}

@article{Anegawa:2025tio,
    author = "Anegawa, Takanori and Tamaoka, Kotaro",
    title = "{Holographic Absolutely Maximally Entangled States in Black Hole Interiors}",
    eprint = "2508.07634",
    archivePrefix = "arXiv",
    primaryClass = "hep-th",
    doi = "10.1103/w7nc-l9s5",
    journal = "Phys. Rev. Lett.",
    volume = "135",
    number = "26",
    pages = "261601",
    year = "2025"
}

\clearpage
\onecolumngrid
\setcounter{section}{0}
\setcounter{subsection}{0}
\setcounter{equation}{0}
\setcounter{figure}{0}
\setcounter{table}{0}
\setcounter{secnumdepth}{2}
\renewcommand{\thesection}{\Roman{section}}
\renewcommand{\thesubsection}{\Alph{subsection}}
\renewcommand{\theequation}{S\arabic{equation}}
\renewcommand{\thefigure}{S\arabic{figure}}
\renewcommand{\thetable}{S\arabic{table}}
\providecommand{\theHsection}{}
\providecommand{\theHsubsection}{}
\providecommand{\theHequation}{}
\providecommand{\theHfigure}{}
\providecommand{\theHtable}{}
\renewcommand{\theHsection}{supp.\arabic{section}}
\renewcommand{\theHsubsection}{supp.\arabic{section}.\arabic{subsection}}
\renewcommand{\theHequation}{supp.\arabic{equation}}
\renewcommand{\theHfigure}{supp.\arabic{figure}}
\renewcommand{\theHtable}{supp.\arabic{table}}
\providecommand{\Tr}{\operatorname{Tr}}
\providecommand{\GM}{\mathrm{GM}}
\providecommand{\arcsinh}{\operatorname{arcsinh}}
\providecommand{\arccosh}{\operatorname{arccosh}}
\allowdisplaybreaks[1]
\begin{center}
{\large\bfseries Supplemental Material for\\[3pt]
Growing Einstein--Rosen Bridge with Multi-partite Entanglement}
\end{center}
\vspace{5pt}

\section{Three-qubit example}
\label{sec:sm-qubits}

\subsection{Spectra and local-unitary equivalence}
For the normalized family in the main body,
\begin{equation}
 \ket{\psi(\theta)}=\cos\theta\,\ket{G}+\sin\theta\,\ket{W},
 \qquad
 \ket{G}=\frac{\ket{000}+\sqrt2\ket{111}}{\sqrt3},
 \qquad
 \ket{W}=\frac{\ket{001}+\ket{010}+\ket{100}}{\sqrt3},
 \label{eq:sm-family}
\end{equation}
permutation symmetry makes the three one-qubit density matrices identical.
Tracing out two qubits gives
\begin{equation}
 \rho_A(\theta)=\rho_B(\theta)=\rho_C(\theta)
 =\frac13
 \begin{pmatrix}
 1+\sin^2\theta & \sin\theta\cos\theta\\
 \sin\theta\cos\theta & 2-\sin^2\theta
 \end{pmatrix},
 \qquad \Tr\rho_A=1,\qquad \det\rho_A=\frac29.
 \label{eq:sm-rho}
\end{equation}
The eigenvalues are therefore $\{2/3,1/3\}$, independently of $\theta$.
Purity of the full state fixes the spectrum of each two-qubit density
matrix to $\{2/3,1/3,0,0\}$. Consequently,
\begin{equation}
 S_n(A)=S_n(B)=S_n(C)=S_n(AB)=S_n(AC)=S_n(BC)
 =\frac{1}{1-n}\log\frac{2^n+1}{3^n}.
 \label{eq:sm-spectra}
\end{equation}
Every mutual information $I_n(X:Y)$ equals the same expression.
Its $n\to1$ limit is $\log3-(2/3)\log2$, and its $n=2$ value is
$\log(9/5)$. The density matrices themselves can vary while these spectra
remain fixed.

The local-unitary equivalence stated in the main body follows from
\begin{equation}
 U=\frac1{\sqrt3}\begin{pmatrix}1&-\sqrt2\\-\sqrt2&-1\end{pmatrix},\label{eq:sm-U}
\end{equation}
\begin{equation}
 U^{\otimes3}\ket{G}=-\frac{\ket{G}+\sqrt2\ket{W}}{\sqrt3},
 \qquad
 U^{\otimes3}\ket{W}=-\frac{\sqrt2\ket{G}-\ket{W}}{\sqrt3},
 \qquad
 U^{\otimes3}\ket{\psi(\theta)}=-\ket{\psi(\theta_0-\theta)},
 \label{eq:sm-LU}
\end{equation}
where $\theta_0=\arctan\sqrt2$. In particular, $U^{\otimes3}\ket{\psi(\theta_0)}=-\ket{G}$.

If desired, the $\theta$ dependence can be interpreted as time evolution generated by the Hamiltonian
\begin{equation}
 H=i\omega\bigl(\ket{W}\!\bra{G}-\ket{G}\!\bra{W}\bigr).
 \label{eq:H}
\end{equation}
In particular, $e^{-iHt}\ket{G}=\cos(\omega t)\ket{G}
+\sin(\omega t)\ket{W}$, so that $\theta=\omega t$.
\subsection{Four-copy contraction and monotonicity}
\begin{figure}[t]
 \includegraphics[width=0.5\textwidth]{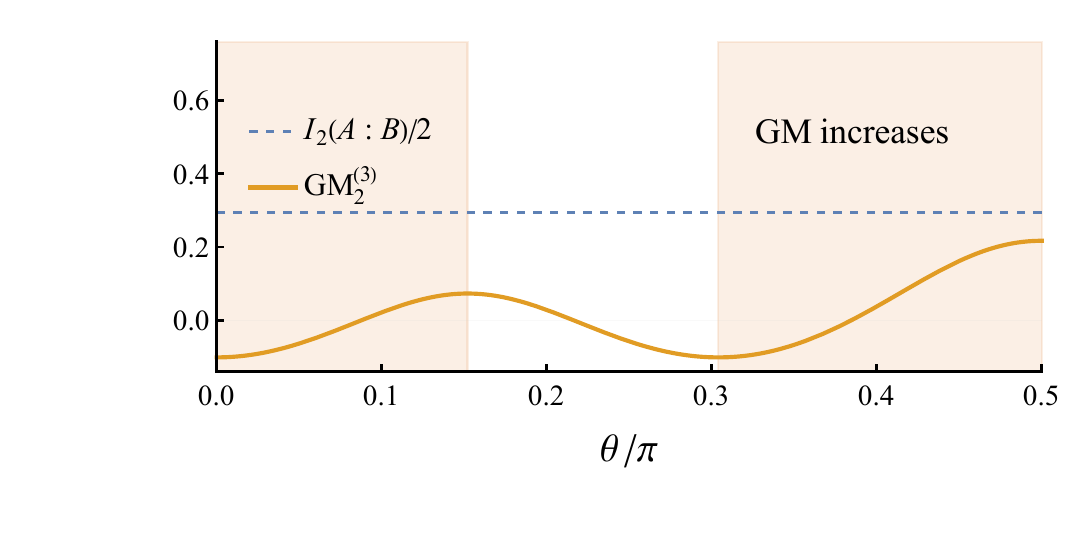}
 \caption{For the shaded region, $\textrm{GM}^{(3)}_2$ increases monotonically.}
 \label{fig:mono}
\end{figure}

We use the normalization of the main body, $S_2^{(3)}=-\tfrac12\log Z_2^{(3)}$.
For a normalized pure state, the four-copy contraction can be evaluated by
realigning $\rho_{AB}$:
\begin{equation}
 R_{aa',bb'}=(\rho_{AB})_{ab,a'b'},
 \qquad
 Z_2^{(3)}=\Tr\!\left[(RR^\dagger)^2\right].
 \label{eq:sm-realignment}
\end{equation}
Indeed, tracing each ket--bra pair over $C$ first gives four copies of
$\rho_{AB}$; the two independent replica swaps on $A$ and $B$ join their
indices into the trace in Eq.~\eqref{eq:sm-realignment}.
To display this contraction compactly, set $x=\tan\theta$ for
$0\leq\theta<\pi/2$. With row and column order $(00,01,10,11)$,
\begin{equation}
 R=\frac{1}{3(1+x^2)}
 \begin{pmatrix}
 1+x^2&x&x&x^2\\
 x&\sqrt2x&x^2&0\\
 x&x^2&\sqrt2x&0\\
 x^2&0&0&2
 \end{pmatrix}.
 \label{eq:sm-R}
\end{equation}
Squaring $RR^\dagger$ and taking its trace yields
\begin{align}
 Z_2^{(3)}(\theta)&=\frac{P(x)}{81(1+x^2)^4},
 \label{eq:sm-polynomial}\\
 \GM_2^{(3)}(\theta)&=-\frac12\log\frac{P(x)}{81(1+x^2)^4}
 -\frac32\log\frac95,
 \label{eq:sm-gm-qubits}
\end{align}
where
\begin{align}
P(x)&=17+12x^2+8\sqrt2x^3+90x^4+24\sqrt2x^5+76x^6+9x^8.
\end{align}
See FIG.~\ref{fig:mono} for the plot.

\section{Gravity calculation}
\label{sec:sm-gravity}

\subsection{Embedding coordinates and mutual information}
We use the BTZ metric and time convention of the main body, with AdS radius
one, $\phi\sim\phi+2\pi$, $\beta=2\pi/r_+$, and
$t_A=-t+i\beta/2$, $t_B=t$. Both intervals have length $L$, and $C$
is the union of their complementary intervals. We take $\pi<L<2\pi$
and $L,2\pi-L\gg\beta$. A common radial cutoff is denoted by $r_b$.
All cutoff-suppressed terms are dropped before expanding at large $r_+$.

We first express each bulk/boundary point in terms of embedding coordinates in $\mathbb R^{2,2}$. The inner product and right-exterior embedding are then given by
\begin{align}
 X\cdot Y&=-X^0Y^0-X^1Y^1+X^2Y^2+X^3Y^3,
 \qquad X\cdot X=-1,\nonumber\\
 X(t,r,\phi)&=\frac1{r_+}\left(
 r\cosh(r_+\phi),\sqrt{r^2-r_+^2}\sinh(r_+t),
 r\sinh(r_+\phi),\sqrt{r^2-r_+^2}\cosh(r_+t)\right).
 \label{eq:sm-embedding}
\end{align}
The boundary limit is $X=(r_b/r_+)P+\cdots$, where $P^2=0$.
The four endpoint vectors are
\begin{equation}
 \begin{aligned}
 P_{A\pm}&=\left(\cosh\frac{r_+L}{2},\sinh(r_+t),
 \pm\sinh\frac{r_+L}{2},-\cosh(r_+t)\right),\\
 P_{B\pm}&=\left(\cosh\frac{r_+L}{2},\sinh(r_+t),
 \pm\sinh\frac{r_+L}{2},+\cosh(r_+t)\right).
 \end{aligned}
 \label{eq:sm-endpoint-vectors}
\end{equation}
The geodesic lengths needed below follow directly from their inner products:
\begin{equation}
 \sigma_{bb}(X,Y)=\arccosh(-X\cdot Y),\qquad
 \sigma_{\partial b}(P,X)=\log(-2P\cdot X)+\log\frac{r_b}{r_+},\qquad
 \sigma_{\partial\partial}(P,Q)=\log(-2P\cdot Q)+2\log\frac{r_b}{r_+}.
 \label{eq:sm-distances}
\end{equation}
In particular, a same-boundary cap of angular size $L$ has length
$2\log[(2r_b/r_+)\sinh(r_+L/2)]$, while an aligned cross-boundary
geodesic has length $2\log[(2r_b/r_+)\cosh(r_+t)]$.
The HRT comparison therefore gives~\cite{Goto:2021gve}
\begin{align}
 S(A)=S(B)&=\frac1{2G_N}\log\!\left[\frac{2r_b}{r_+}\sinh\frac{r_+L}{2}\right],
 \label{eq:sm-SA}\\
 S(C)=S(AB)&=\frac1{G_N}\log\frac{2r_b}{r_+}
 +\frac1{G_N}\min\left\{\log\cosh(r_+t),\,
 \log\sinh\frac{r_+(2\pi-L)}2\right\}.
 \label{eq:sm-SC}
\end{align}
For $S(A)$ and $S(B)$, a complement cap plus a horizon has leading
length $r_+(4\pi-L)>r_+L$, so it does not dominate. For $S(AB)$,
two interval caps are longer than two complement caps because $L>\pi$.
Equating the two entries in Eq.~\eqref{eq:sm-SC} yields
\begin{equation}
 t_{\rm MI}^{\rm exact}=\frac1{r_+}\arccosh\!\left[
 \sinh\frac{r_+(2\pi-L)}2\right],\qquad
 t_{\rm MI}=\pi-\frac L2\quad\text{at leading order}.
 \label{eq:sm-hrt-time}
\end{equation}
Using $s_{\rm BH}=r_+/(4G_N)$, as in the main body, we obtain
\begin{equation}
 \frac12 I(A:B)=s_{\rm BH}\max\{L-2t,\,2L-2\pi\}+O(r_+^0),
 \qquad I(A:C)=I(B:C)=S(C).
 \label{eq:sm-mi}
\end{equation}
The second equality uses purity and $S(A)=S(B)$.
Together with $S(AC)=S(B)$ and $S(BC)=S(A)$, this shows that all
subsystem entropies and mutual informations are stationary
after the HRT exchange.

\subsection{Direct extremization of the connected network (Phase I and II)}
For the connected network, the two upper endpoints join at $X$, the
two lower endpoints join at $Y$, and one internal branch connects $X$
and $Y$. We evaluate its holographic multi-entropy~\cite{Gadde:2022cqi}
by directly extremizing the total length
\begin{equation}
 \mathcal L_{\rm conn}(X,Y)=
 \sigma_{\partial b}(P_{A+},X)+\sigma_{\partial b}(P_{B+},X)
 +\sigma_{\partial b}(P_{A-},Y)+\sigma_{\partial b}(P_{B-},Y)
 +\sigma_{bb}(X,Y).
 \label{eq:sm-network-functional}
\end{equation}
This corresponds to phase I and II. 
By symmetry, $r_X=r_Y=r_0$, $\phi_X=-\phi_Y=\phi_0$.
For $t>0$, the time-reflection symmetry fixes $t_X=t_Y=t_0=i\beta/4$.
On the future-interior branch, the corresponding real embedding vectors are
\begin{equation}
 X=\frac1{r_+}\left(r_0\cosh(r_+\phi_0),\sqrt{r_+^2-r_0^2},
 r_0\sinh(r_+\phi_0),0\right),\qquad
 Y=\frac1{r_+}\left(r_0\cosh(r_+\phi_0),\sqrt{r_+^2-r_0^2},
 -r_0\sinh(r_+\phi_0),0\right).
 \label{eq:sm-XY}
\end{equation}
Taking the inner products gives
\begin{align}
 -P_{A+}\cdot X=-P_{B+}\cdot X
 =-P_{A-}\cdot Y=-P_{B-}\cdot Y
 &=\frac{r_0}{r_+}\cosh\!\left[r_+(L/2-\phi_0)\right]
 +\sqrt{1-\frac{r_0^2}{r_+^2}}\sinh(r_+t),
 \label{eq:sm-boundary-products}\\
 -X\cdot Y&=1+2\frac{r_0^2}{r_+^2}\sinh^2(r_+\phi_0).
 \label{eq:sm-bulk-product}
\end{align}
Consequently, the exact reduced length is
\begin{align}
 \mathcal L_{\rm conn}(r_0,\phi_0)
 ={}&4\log\frac{r_b}{r_+}
 +4\log\!\left[2\left\{
 \frac{r_0}{r_+}\cosh\!\left[r_+(L/2-\phi_0)\right]
 +\sqrt{1-\frac{r_0^2}{r_+^2}}\sinh(r_+t)\right\}\right]\nonumber\\
 &+\arccosh\!\left[1+2\frac{r_0^2}{r_+^2}\sinh^2(r_+\phi_0)\right].
 \label{eq:sm-exact-functional}
\end{align}

We now hold $0<t<L/2$ fixed and take $r_+\to\infty$.
The stationary point has $\phi_0=L/2-t+\mathcal{O}(1/r_+)$, so both
$r_+\phi_0$ and $r_+(L/2-\phi_0)$ are large. Expanding
Eq.~\eqref{eq:sm-exact-functional} while retaining its finite terms gives
\begin{align}
 \mathcal L_{\rm conn}
 ={}&4\log\frac{r_b}{r_+}
 +4\log\!\left[\frac{r_0}{r_+}e^{r_+(L/2-\phi_0)}
 +\sqrt{1-\frac{r_0^2}{r_+^2}}e^{r_+t}\right]
 +2r_+\phi_0+2\log\frac{r_0}{r_+}.
 \label{eq:sm-expanded-functional}
\end{align}
In particular, the internal length is
$2r_+\phi_0+2\log(r_0/r_+)$.

The angular variation gives
\begin{equation}
 0=\frac{\partial\mathcal L_{\rm conn}}{\partial\phi_0}
 =-4r_+\frac{(r_0/r_+)e^{r_+(L/2-\phi_0)}}
 {(r_0/r_+)e^{r_+(L/2-\phi_0)}+\sqrt{1-r_0^2/r_+^2}\,e^{r_+t}}
 +2r_+.
 \label{eq:sm-angular-variation}
\end{equation}
It sets the two terms in the denominator equal and hence gives
\begin{equation}
 \frac{r_0}{r_+}e^{r_+(L/2-\phi_0)}
 =\sqrt{1-\frac{r_0^2}{r_+^2}}\,e^{r_+t},\qquad
 \phi_0=\frac L2-t-\frac1{r_+}\log\sqrt{\frac{r_+^2}{r_0^2}-1}.
 \label{eq:sm-angular-saddle}
\end{equation}
Eliminating $\phi_0$ in Eq.~\eqref{eq:sm-expanded-functional} then yields
\begin{equation}
 \mathcal L_{\rm conn}
 =4\log\frac{r_b}{r_+}+r_+(L+2t)+4\log2
 +\log\!\left(1-\frac{r_0^2}{r_+^2}\right)
 +4\log\frac{r_0}{r_+}.
 \label{eq:sm-radial-functional}
\end{equation}
Since the angular derivative already vanishes, radial stationarity reduces to
\begin{equation}
 \frac4{r_0}-\frac{2r_0}{r_+^2-r_0^2}=0,
 \qquad
 r_0=\sqrt{\frac23}\,r_++\mathcal{O}(r_+^0),\qquad
 \phi_0=\frac L2-t+\frac{\log2}{2r_+}.
 \label{eq:sm-connected-saddle}
\end{equation}
Thus the two junctions lie inside the horizons. Substitution in
Eq.~\eqref{eq:sm-radial-functional} gives
\begin{equation}
 \mathcal L_{\rm conn}
 =4\log\frac{r_b}{r_+}+r_+(L+2t)+6\log\frac2{\sqrt3},
 \qquad
 S_{\rm conn}^{(3)}=\frac1{G_N}\log\frac{r_b}{r_+}
 +s_{\rm BH}(L+2t)+2\kappa_3,
 \label{eq:sm-connected-result}
\end{equation}
where $\kappa_3=\frac{3}{4G_N}\log(2/\sqrt3)$ is the constant in the main body.
The separate branches evaluate to
\begin{equation}
 \sigma_{\partial b}(P_{A+},X_*)
 =\log\frac{r_b}{r_+}+r_+t+\log\frac2{\sqrt3},
 \qquad
 \sigma_{bb}(X_*,Y_*)=r_+(L-2t)+2\log\frac2{\sqrt3}.
 \label{eq:sm-branch-lengths}
\end{equation}
All four external legs have the first length. Their $4r_+t$ growth and
the internal branch's $-2r_+t$ contribution give the net $2r_+t$ growth
of the connected network. The errors above are exponentially small for
fixed $t>0$ and $L/2-t>0$.

At $t=0$, one instead evaluates the length directly on the bifurcation
slice, $r_0=r_+$. Equation~\eqref{eq:sm-exact-functional} reduces to
$4\log(r_b/r_+)+4\log\{2\cosh[r_+(L/2-\phi_0)]\}+2r_+\phi_0$.
Its angular derivative gives
$\phi_0=L/2-\log\sqrt3/r_+$. The HRT subtraction then yields
\begin{equation}
 \GM^{(3)}(0)=2\kappa_3
 \label{eq:sm-initial-value}
\end{equation}
This has no term of order $r_+$, consistently with phase I of the main body.

\subsection{Late  time candidates (Phase III and IV)}
The two interval caps have total length
\begin{equation}
 \mathcal L_{\rm caps}
 =4\log\!\left[\frac{2r_b}{r_+}\sinh\frac{r_+L}{2}\right]
 =4\log\frac{r_b}{r_+}+2r_+L+\mathcal{O}(r_+^0).
 \label{eq:sm-caps}
\end{equation}
This gives the result for phase III. 
There is also a distinct winding candidate. For example, attach the two
endpoints of $B$ to vertices $X,Y$ on the right exterior, with
$r_X=r_Y=r_0>r_+$, $t_X=t_Y=t$, and $\phi_X=-\phi_Y=\phi_0$.
Join the vertices by a direct branch and a branch in the adjacent winding
class. The opposite boundary contributes one cap over the complement of
$A$, whose angular size is $2\pi-L$. 

The direct and winding branches use angular separations $2\phi_0$ and
$2\pi-2\phi_0$, respectively. Substituting the exterior embedding into
Eq.~\eqref{eq:sm-distances} gives the loop length, including its two
external legs,
\begin{align}
 \mathcal L_{\rm loop}(r_0,\phi_0)
 ={}&2\log\frac{r_b}{r_+}
 +2\log\!\left[2\left\{
 \frac{r_0}{r_+}\cosh\!\left[r_+(L/2-\phi_0)\right]
 -\sqrt{\frac{r_0^2}{r_+^2}-1}\right\}\right]\nonumber\\
 &+\arccosh\!\left[1+2\frac{r_0^2}{r_+^2}\sinh^2(r_+\phi_0)\right]
 +\arccosh\!\left[1+2\frac{r_0^2}{r_+^2}
 \sinh^2\!\left[r_+(\pi-\phi_0)\right]\right].
 \label{eq:sm-loop-functional}
\end{align}
At large $r_+$, the last two terms sum to
$2\pi r_++4\log(r_0/r_+)$. Angular stationarity therefore gives
$\phi_0=L/2+o(1/r_+)$, and
\begin{equation}
 \mathcal L_{\rm loop}
 =2\log\frac{r_b}{r_+}+2\pi r_++2\log2
 +2\log\!\left(\frac{r_0}{r_+}-\sqrt{\frac{r_0^2}{r_+^2}-1}\right)
 +4\log\frac{r_0}{r_+}.
 \label{eq:sm-loop-expanded}
\end{equation}
Taking its radial derivative gives
\begin{equation}
 -\frac2{\sqrt{r_0^2-r_+^2}}+\frac4{r_0}=0,
 \qquad r_0=\frac2{\sqrt3}r_+,
 \label{eq:sm-loop-saddle}
\end{equation}
so these vertices are outside the horizon. Therefore, we have
\begin{align}
 \mathcal L_{\rm loop}
 &=2\log\frac{r_b}{r_+}+2\pi r_++6\log\frac2{\sqrt3}
 \label{eq:sm-loop-result}\\
 \mathcal L_{\rm wind}
 &=\mathcal L_{\rm loop}
 +2\log\!\left[\frac{2r_b}{r_+}\sinh\frac{r_+(2\pi-L)}2\right]
 \nonumber\\
 &=4\log\frac{r_b}{r_+}+r_+(4\pi-L)+6\log\frac2{\sqrt3}.
 \label{eq:sm-winding-result}
\end{align}
Thus $S_{\rm wind}^{(3)}=\mathcal L_{\rm wind}/(4G_N)$ contains the
same $2\kappa_3$ term as the connected candidate. The loop and opposite
complement cap together form the required partition; the two internal
branches form its noncontractible cycle, and each boundary entangling
point emits exactly one wall. Shortening the connected network's internal
branch by itself would not give this topology.

\end{document}